\documentclass[aps,prl,twocolumn,superscriptaddress]{revtex4}
\usepackage{mathrsfs}
\usepackage{epsfig}
\usepackage{graphicx}
\usepackage{amsfonts}
\usepackage[figuresright]{rotating}
\usepackage{amssymb}
\usepackage{amsmath}
\usepackage{dcolumn}
\usepackage{bm}
\usepackage{color}

\def\be{\begin{equation}} \def\ee{\end{equation}}
\def\bea{\begin{eqnarray}} \def\eea{\end{eqnarray}}

\newcommand{\WQCASQC} {Wilczek Quantum Center and Key Laboratory of Artificial Structures and Quantum Control, School of Physics and Astronomy, Shanghai Jiao Tong University, Shanghai 200240, China}

\newcommand{\BHU}{
School of Physics, Key Laboratory of Micro-Nano Measurement-Manipulation
and Physics (Ministry of Education), Beihang University, Beijing 100191, China }

\newcommand{\SRCQC}{Shanghai Research Center for Quantum Sciences, Shanghai 201315, China}

\newcommand{\zhiyuan} {Zhiyuan College, Shanghai Jiao Tong University, Shanghai 200240, China}

\begin{document}
\title{Symmetry protected topological edge modes and emergent partial time reversal symmetry breaking in open quantum many-body systems}

\author{Zijian Wang}
\thanks{These authors contributed equally to this work.}
\affiliation{\WQCASQC}
\affiliation{\zhiyuan}

\author{Qiaoyi Li}
\thanks{These authors contributed equally to this work.}
\affiliation{\BHU}

\author{Wei Li}
\email{w.li@buaa.edu.cn}
\affiliation{\BHU}

\author{Zi Cai}
\email{zcai@sjtu.edu.cn}
\affiliation{\WQCASQC}
\affiliation{\SRCQC}

\begin{abstract}  
Symmetry-protected topological edge modes are one of the most remarkable phenomena in topological physics. Here, we formulate and quantitatively examine the effect of a quantum bath on these topological edge modes. Using the  density matrix renormalization group method, we study the ground state of a composite system of spin-1 quantum chain, where the system and the bath degrees of freedom are treated on the same footing. We focus on the dependence of these edge modes on the global/partial symmetries of system-bath coupling and on the features of the quantum bath.  It is shown that the time-reversal symmetry(TRS) plays a special role for an open quantum system, where an emergent partial TRS breaking will result in a TRS-protected topological mode diffusing from the system edge into the bath, thus make it useless for quantum computation.
\end{abstract}


\maketitle

{\it Introduction --} Topological phases are characterized by robust topological features rather than by  symmetries and their spontaneous breaking. More often than not, these topological features are only robust against perturbations preserving certain symmetries. For instance, the helical edge modes in the $Z_2$ topological insulator are protected by time-reversal symmetry (TRS)\cite{Kane2005,Fu2007}. Such symmetry-protected phenomena are not only restricted to free fermions, but can also appear in interacting systems\cite{Chen2011,Fidkowski2011,Schuch2011,Turner2011,Pollmann2012,Magnifico2019}, such as the Haldane chain\cite{Haldane1983}. Such topological quantum phases, dubbed  ``symmetry protected topological''(SPT) phases\cite{Gu2009,Chen2012,Chen2013},  are bulk-gapped phases with gapless or degenerate edge modes protected by certain global symmetries. These degenerate edge modes span a subspace  immune to  symmetry-preserved perturbations, thereby exhibit potential applications in quantum computation and information processing\cite{Alicea2012,Nayak2008}.

Almost all realistic quantum systems are coupled to its surroundings and hence an open quantum system.  Understanding the SPT phase in the presence of a bath is not only of practical importance in quantum information, but is also of fundamental significance from the perspective of topological physics. For instance, an SPT phase is usually defined for a pure state (the groundstate).  However, it is unclear to what extent  this topological phase can be generalized to open systems which are generally characterized by mixed states\cite{Uhlmann1986,Viyuela2014a,Huang2014,Budich2015b,Bardyn2018}. Furthermore, symmetry is one of the most important ingredients of SPT phases, however, its definition in an open system is more subtle than its counterparts in isolated systems, because it involves not only the system but also the bath and  system-bath (SB) coupling. As a consequence, one needs to distinguish the global symmetries with respect to all(system+bath) variables  from the partial symmetries with respect to the system variables only. From a practical perspective, the partial symmetries are more relevant to the experimental observable because of a lack of bath information in most open quantum systems, especially those in the solid state setups. The global symmetry, on the other hand, is important in the sense that it contains the complete information of an open system.  Such a subtlety of symmetry in open quantum systems might give rise to intriguing SPT physics absent in closed systems.


 \begin{figure}[htb]
\includegraphics[width=0.99\linewidth]{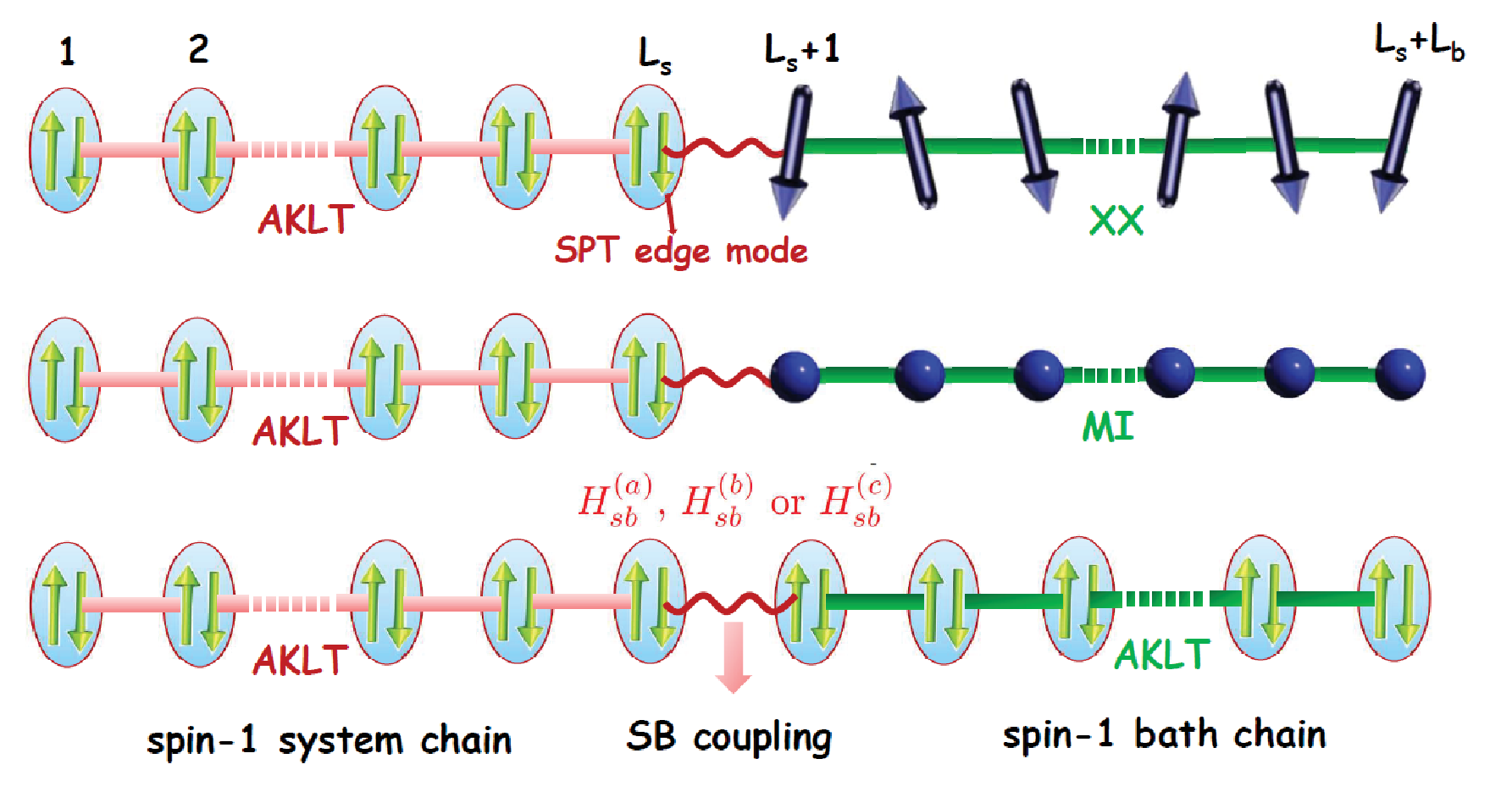}
\caption{(Color online) Schematic  of  the symmetry-protected topological systems (AKLT state) coupled to different quantum baths via various SB couplings.
} \label{fig:fig1}
\end{figure}

In this study, we explore the SPT edge mode in the presence of a quantum bath by considering a composite quantum system, which can be split into an SPT ``system'' and a quantum ``bath'', thus allows us to examine the roles of partial and global symmetries respectively.  We focus on the dependence of the SPT edge mode on the symmetry of the SB coupling as well as the feature of the quantum bath. It is shown that the time-reversal symmetry (TRS) plays a special role in open quantum systems: even though the microscopic Hamiltonian of the composite system (system+bath) preserves both the global and partial TRS, the latter can be spontaneously broken which gives rises to remarkable consequence that the TRS-protected topological mode will diffuse from the system edge into the bath, thus is useless for quantum computation. In a seminar work\cite{McGinley2020}, it was proposed by McGinley {\it et al} that such an emergent partial TRS breaking is responsible for the decoherence of a TRS-protected edge mode in an open systems which preserve both the global and partial TRS. Different from Ref.\cite{McGinley2020}, here we focus on the groundstate properties of the total system rather than its time evolution. In addition, the ``system'' considered here is a genuine interacting quantum many-body system, whose SPT is not only protected by TRS, but also by other symmetries. Furthermore, it is shown that the quantum bath also plays an important role in determining the properties of the SPT edge modes.

{\it Model and method --}  To study the effect of various quantum baths on the SPT phase,  the bath degrees of freedom were treated on the same footing as the system variables, instead of tracing them out, as is traditionally done in open quantum systems\cite{Breuer2002}.   To this end,  we consider a composite system of an one-dimensional(1D) $S=1$ quantum spin model as shown in Fig.\ref{fig:fig1}, whose Hamiltonian (Ham) can be divided into three parts:
\begin{equation}
H_{tot}=H_s+H_{sb}+H_b \label{eq:Hamt}
\end{equation}
where $H_s$ is the system Ham with an SPT ground state, $H_b$ is the bath Ham and $H_{sb}$ is coupling between them.

Throughout this study, we choose $H_s$ as the Affleck-Kennedy-Lieb-Tasaki (AKLT) Hamiltonian\cite{Affleck1987}:
\begin{small}
\begin{equation}
H_s=\sum_{i=1}^{L_s-1} J[\mathbf{S}_i\cdot \mathbf{S}_{i+1}+\frac 13 (\mathbf{S}_i\cdot \mathbf{S}_{i+1})^2]\label{eq:AKLT}
\end{equation}
\end{small}
$S^{x,y,z}_i$  are the generators of the SO(3) group in the spin-1 representation\cite{Supplementary}, $L_s$ is the length of the system chain.  The ground state of Ham.(\ref{eq:AKLT}) with open boundary condition supports two edge modes in its ends, each of which carries a fractionalized spin-$\frac 12$ that can point along any direction, and hence contributes a two-fold degeneracy to ground state. This state is a prototypical example of a non-trivial SPT phase with simple symmetries\cite{Gu2009,Berg2009,Pollmann2012}:  TRS, SO(3) spin rotational symmetry (SRS) and discrete $Z_2\times Z_2$ SRS (i.e. $\mathcal{R}^x\mathcal{R}^y$ with $\mathcal{R}^{x,y,z}$ being the $\pi$-rotation along the $x,y,z$ axes). Each of these symmetries can protect the edge modes and the corresponding degeneracy.  The bath is also chosen as a spin-1 quantum chain, whose Ham $H_b$ can take different forms as stated latter on.

SB coupling only occurs on the bond $[L_s L_s+1]$ connecting  the system and bath chain, which takes the general form: $H_{sb}=J'\sum_\alpha A_{L_s}^\alpha\otimes B_{L_s+1}^\alpha$ where $A^\alpha$/$B^\alpha$ is the  spin operator of the  system/bath, and is assumed to be hermite for simplicity.  $\alpha$ is the index of coupling channel and $J'$ denotes the coupling strength. Throughout the paper, we focus on the cases with SB couplings preserving the global TRS and $Z_2\times Z_2$ discrete SRS, and examine the role of the partial symmetries. To this end, we consider three SB couplings with different partial symmetries:
\begin{small}
\begin{eqnarray}
H_{sb}^{(a)}&=&J'(S_{L_s}^x \otimes S_{L_s+1}^x+S_{L_s}^y\otimes  S_{L_s+1}^y)  \label{eq:Hama}\\
H_{sb}^{(b)}&=&J'(S_{L_s}^{xz}\otimes S_{L_s+1}^{xz}+S_{L_s}^{xy}\otimes S_{L_s+1}^{xy}+S_{L_s}^{yz}\otimes S_{L_s+1}^{yz}) \label{eq:Hamb}\\
H_{sb}^{(c)}&=&J'(S_{L_s}^{xx}\otimes S_{L_s+1}^{xx}+S_{L_s}^{yy}\otimes S_{L_s+1}^{yy}) \label{eq:Hamc}
\end{eqnarray}
\end{small}
where $S^{\gamma\beta}=\eta_{\gamma\beta}\{S^\gamma,S^\beta\}$ with $\eta_{\gamma\beta}=(2-\delta_{\gamma\beta})/2$.  Since the global symmetries are preserved from Eq.(\ref{eq:Hama}) to Eq.(\ref{eq:Hamc}), we distinguish these three SB couplings according to the partial symmetry:  $H_{sb}^{(a)}$  breaks both partial TRS and $Z_2\times Z_2$ SRS;   $H_{sb}^{(b)}$ preserves the former but breaks latter, while in $H_{sb}^{(c)}$, both symmetries are preserved. To enumerate all the possibilities, in principle, one needs to consider another case that breaks the partial TRS but preserves the $Z_2\times Z_2$ SRS. However, as we shown in the SM, there is no qualitatively difference between the SPT edge mode in this case and the case with $H_{sb}^{(c)}$.

The ground state of Ham.(\ref{eq:Hamt}) is studied by the Density-Matrix-Renormalization-Group method adapted to the matrix-product-state
formalism with standard two-site variational update\cite{White1992,Schollwock2005,Schollwock2011}, and we retain
up to 300 bond states that leads to negligible truncation errors ($\leq 10^{-10}$) in the calculations. More details of the numerical methods and convergence benchmarks can be found in the SM\cite{Supplementary}. The system chain is assumed to be sufficiently long so that the spin-$\frac 12$ edge modes at its two ends decouple with each other.  A small magnetic field is imposed on last site of the system chain ($H_h=-h_z S_{L_s}^z$) to monitor the behavior of the edge mode coupling to the bath, and the  local magnetization $m_{L_s}=\langle S_{L_s}^z\rangle$ is calculated in the limit of $h_z\rightarrow 0$. Without SB coupling, the edge spin-$\frac 12$ of the AKLT chain can be polarized along z-direction by infinitesimal field, which can be considered as a probe of the local edge mode in our numerical simulation. We then adiabatically turn on the SB coupling, and study its effect on  spin-$\frac 12$ edge mode coupling to the bath.

 \begin{figure*}[htb]
\includegraphics[width=0.32\linewidth,bb=1 1 675 621]{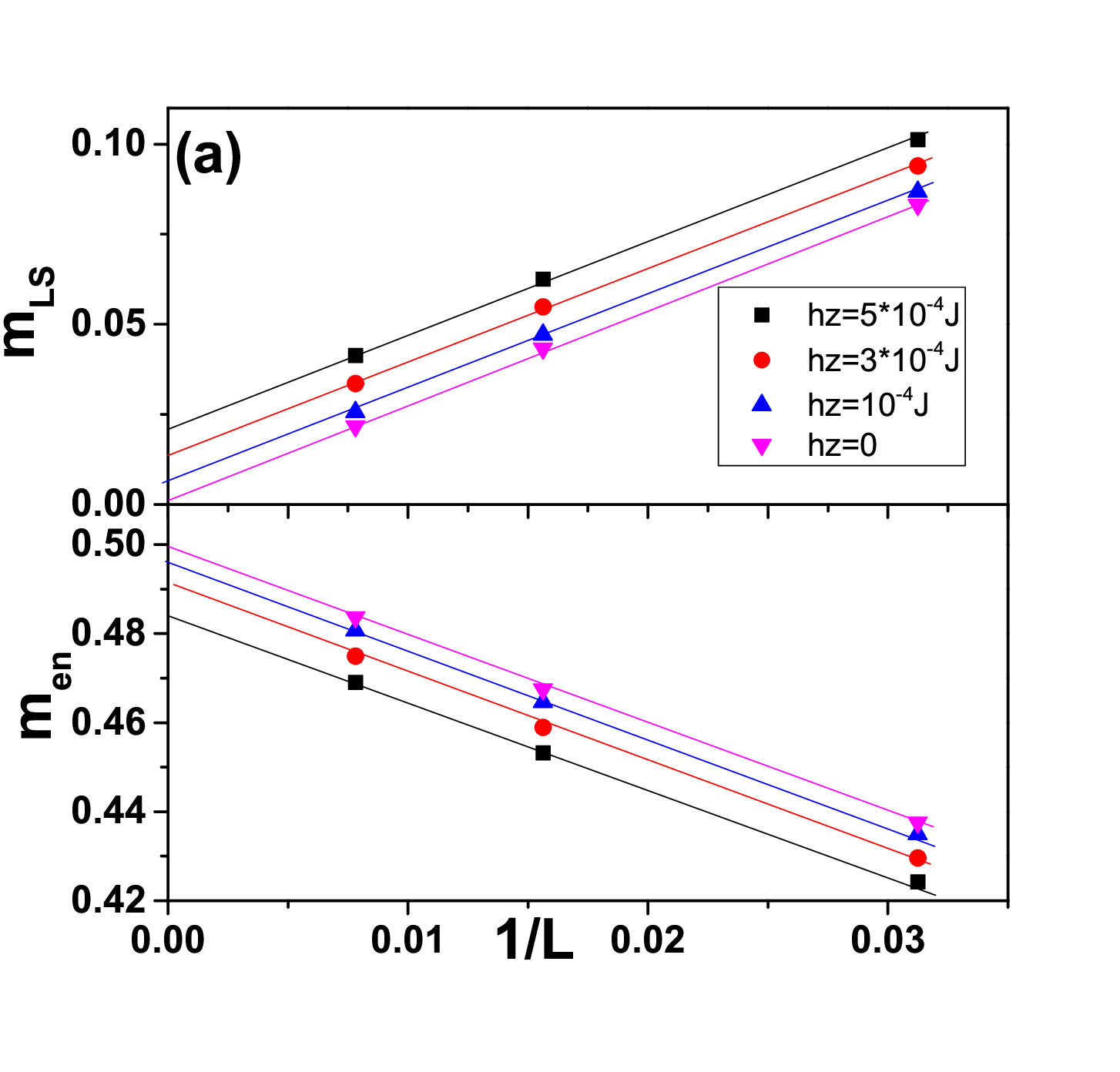}
\includegraphics[width=0.32\linewidth,bb=1 1 675 621]{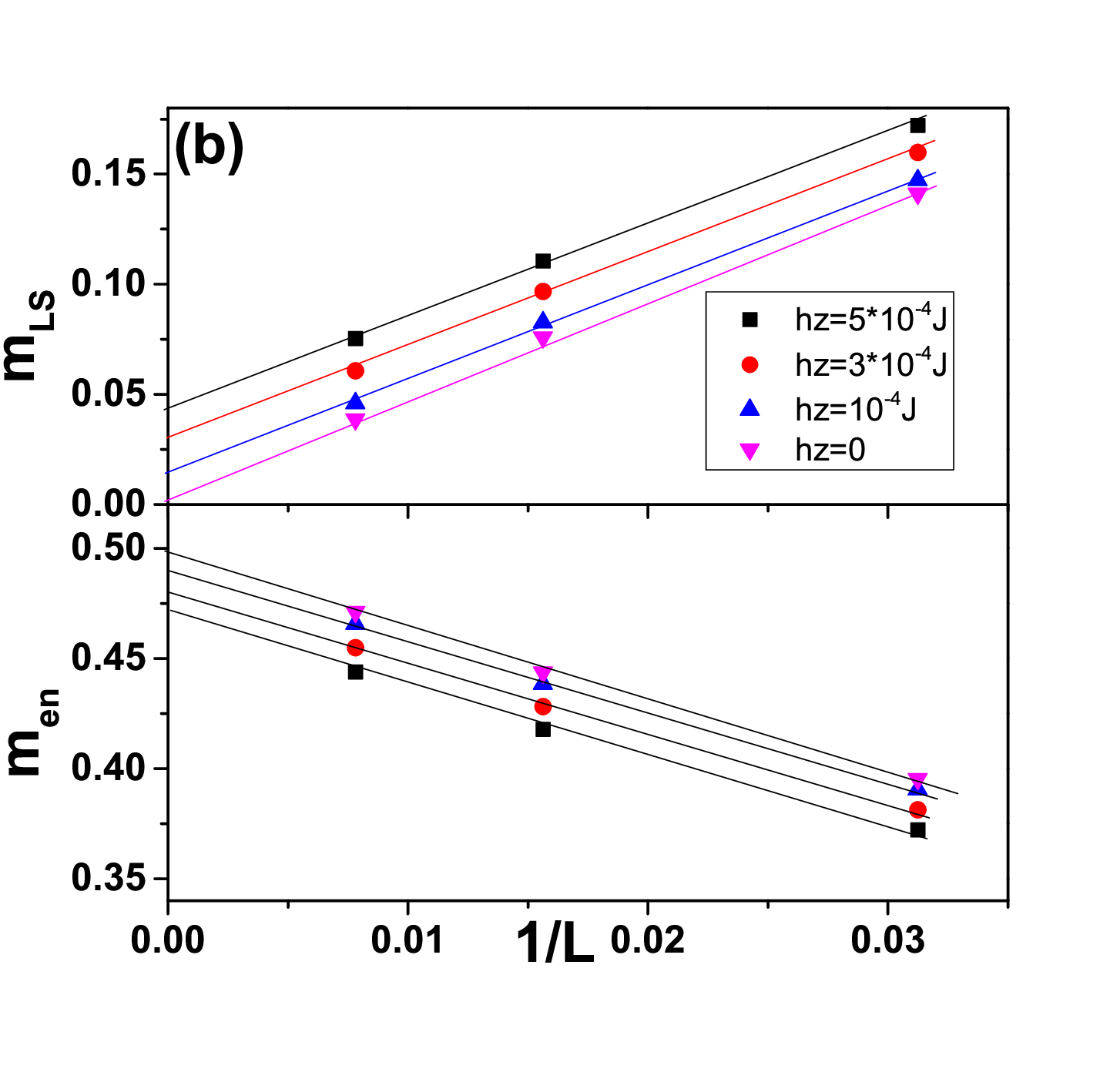}
\includegraphics[width=0.33\linewidth,bb=1 1 686 600]{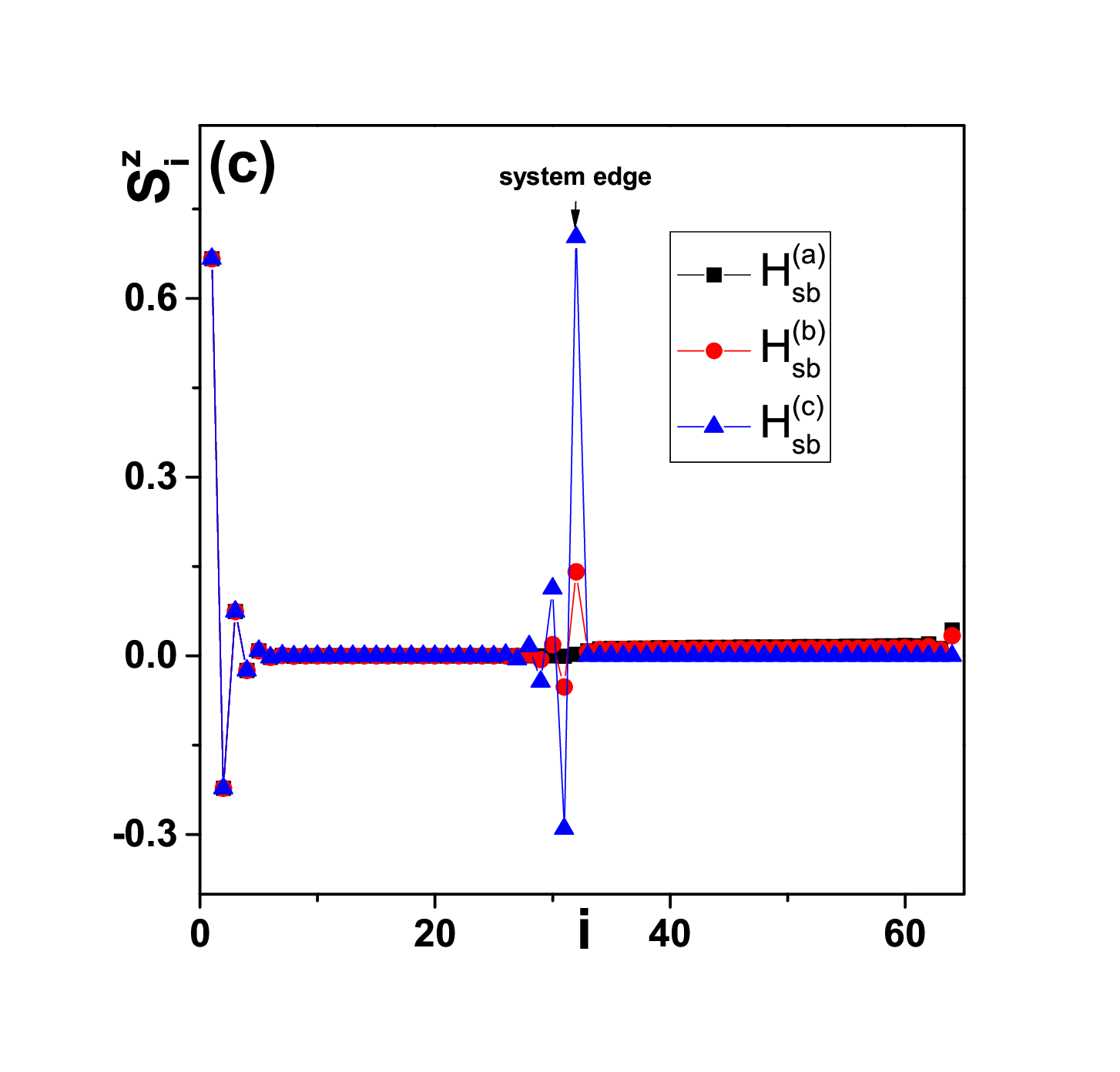}
\caption{(Color online). Finite size scaling of the magnetization on the edge site of the system (upper panels)  and within the bath chain (lower panels) for the AKLT model coupled to the XX model via SB coupling: (a) $H_{sb}^{(a)}$ and  (b) $H_{sb}^{(b)}$ in the presence of various small local magnetic fields $h_z$ (the results of $h_z=0$ are obtained via  $h_z\rightarrow 0$ extrapolations).  (c) Spatial distribution of  magnetization in the ground state of the composite system with $L_s=L_b=32$ and different SB couplings (square for $H_{sb}^{(a)}$, circle for $H_{sb}^{(b)}$, and triangle for $H_{sb}^{(c)}$). $J_b=0.1J$ for (a)-(c) and $J'=0.02J$ for (a) and $J'=0.2J$ for (b) and (c).}\label{fig:fig2}
\end{figure*}
{\it Gapless quantum bath and emergent partial TRS breaking --} As a beginning, we consider the case with gapless quantum bath. Motivated by a recent intriguing
proposal of modeling the bath via an engineered spin chain\cite{Ramos2016,Vermersch2016}, the bath Ham is chosen as a spin-1 XX model:
\begin{equation}
\begin{small}
H_b=J_b\sum_{i=L_s+1}^{L_b+L_s-1} (S_i^xS_{i+1}^x+S_i^y S_{i+1}^y )\label{eq:Hame}
\end{small}
\end{equation}
Ham.(\ref{eq:Hame}) become gapless when its length $L_b\rightarrow \infty$.

First, we consider the case with SB coupling $H_{sb}^{(a)}$ in Eq.(\ref{eq:Hama}), which breaks the partial TRS and $Z_2\times Z_2$ SRS.  Fig.\ref{fig:fig2}(a) shows finite size scaling of the local magnetization $m_{L_s}$ and total magnetization in the bath chain $m_{en}=\sum_{i=L_s+1}^{L_s+L_b} \langle S_i^z\rangle$ for different $h_z$.  In the thermodynamic limit ($L_b\rightarrow\infty$), the spin-$\frac 12$ edge mode vanishes on the last site of the system chain ($m_{L_s}\rightarrow 0$ for $h_z\rightarrow 0^+$). At the same time, the total magnetization in the bath chain $m_{en}$ approach 0.5, which indicates that the spin-$\frac 12$ edge mode originally located on site $L_s$, has diffused into the gapless bath chain.   In other words,  the spin-$\frac 12$ ``edge'' mode in this open SPT system is extensive and no longer located on the system edge,  thus is useless in practice for topological quantum computation.  On the other hand, for the ground state of the composite system, as long as the global TRS is preserved, the degeneracy corresponding to the ``edge'' mode still exists even though the spin-$\frac 12$ edge mode is no longer located on the system edge but becomes an extensive object in the bath.

To examine the role of the TRS, the SB coupling $H_{sb}^{(b)}$ in Eq.(\ref{eq:Hamb}) is considered, the system variables ({\it e.g.} $S_{L_s}^{xy}$) of which preserves the TRS but breaks the  $Z_2\times Z_2$ SRS. From Fig.\ref{fig:fig2} (b), it can be seen  that the results are similar to those for $H_{sb}^{(a)}$, where $m_{L_s}\rightarrow 0$ and  $m_{en}\rightarrow 0.5$ in the limit of $L_b\rightarrow \infty$ and $h_z\rightarrow 0^+$, indicating that the partial TRS of  $H_{sb}^b$  cannot prevent the edge mode from diffusing into the gapless bath. This phenomenon can be understood via a perturbation analysis. For a general form of the SB coupling: $H_{sb}=J'\sum_\alpha A_{L_s}^\alpha\otimes B_{L_s+1}^\alpha$  in the limit of $J'\ll \Delta$ with $\Delta$ is the gap of the SPT phase, one can treat it as a perturbation and derive an effective SB coupling Hamiltonian whose system variables only operate in the SPT subspace spanned by the degenerate spin-$\frac 12$ edge modes. Since all $A_{L_s}^\alpha$ preserve the TRS, the effective Hamiltonian to the 1st order of $J'$ is trivial ($A_{L_s}^\alpha$ itself cannot lift the TRS protected degeneracy). The 2nd order perturbation gives rise to\cite{Mila2011}:
\begin{equation}
\begin{small}
H_{sb}^{(2)}=\mathbb{P} H_{sb} \mathbb{Q}\frac{1}{E_0-\mathbb{Q}H_0\mathbb{Q}}\mathbb{Q}H_{sb}\mathbb{P}
\end{small}
\end{equation}
where $H_0=H_s+H_b$, $E_0$ is its ground state energy. $\mathbb{P}$ is the projection operator onto the SPT subspace of the system, which leaves the bath intact. $\mathbb{Q}=\mathbb{I}-\mathbb{P}$. By making the approximation $\frac{1}{E_0-\mathbb{Q}H_0\mathbb{Q}}\sim -\frac 1\Delta$, we obtain
\begin{equation}
H_{sb}^{(2)}\sim -\frac{J'^2}{\Delta}\sum_{\alpha\beta}\mathbb{P} C^{\alpha\beta}_{L_s}\mathbb{P}\otimes B^\alpha_{L_s+1} B^\beta_{L_s+1} \label{eq:Hameff}
\end{equation}
where the operator $C^{\alpha\beta}_{L_s}=A^\alpha_{L_s}\mathbb{Q}A^\beta_{L_s}$  is in general non-hermitian for $\alpha\neq\beta$ and can be divided into hermite and anti-hermite parts as: $C^{\alpha\beta}_{L_s}=\tilde{C}^{\alpha\beta}_H+i \tilde{C}^{\alpha\beta}_A$ where $\tilde{C}^{\alpha\beta}_H=\frac 12[C^{\alpha\beta}_{L_s}+[C^{\alpha\beta}_{L_s}]^\dag]$, $\tilde{C}^{\alpha\beta}_A=\frac 1{2i}[C^{\alpha\beta}_{L_s}-[C^{\alpha\beta}_{L_s}]^\dag]$ and both of them are hermite operators. $A^\alpha_{L_s}$ preserves the TRS, the same is true for $C^{\alpha\beta}_{L_s}$, indicating $\tilde{C}^{\alpha\beta}_H$ ($\tilde{C}^{\alpha\beta}_A$) is time reversal even(odd). Since $\tilde{C}^{\alpha\beta}_H$  preserves  TRS, it doesn't mix  degenerate SPT modes, thus only $\tilde{C}^{\alpha\beta}_A$  contributes to $H_{sb}^{(2)}$ nontrivially. Ham.(\ref{eq:Hameff}) can be rewritten as:
\begin{small}
\begin{equation}
H_{sb}^{(2)}\sim -\frac{J'^2}{2\Delta}\sum_{\alpha\beta} \tilde{A}^{\alpha\beta}_{L_s}\otimes\tilde{B}^{\alpha\beta}_{L_s+1}\label{eq:Hameff2}
\end{equation}
\end{small}
where $\tilde{A}^{\alpha\beta}_{L_s}=\mathbb{P}\tilde{C}^{\alpha\beta}_A\mathbb{P}$, $\tilde{B}^{\alpha\beta}_{L_s+1}=i[B^\alpha_{L_s+1},B^\beta_{L_s+1}]$ and both of them are hermite operators.

The effective Ham.(\ref{eq:Hameff2}) indicates  an emergent partial TRS breaking: even though in the original SB coupling Ham $H_{sb}^{(b)}$, the system variables ($A^\alpha_{L_s}$) preserve the TRS, this doesn't hold for the effective SB coupling Ham.(\ref{eq:Hameff2}), whose system varibales ($\tilde{A}^{\alpha\beta}_{L_s}$) are time reversal odd, thus result in similar consequence with the case of $H_{sb}^{(a)}$. The explicit form of the effective SB coupling is shown in the SM\cite{Supplementary}. Therefore, the partial TRS alone is not sufficient to protect the SPT edge modes from diffusing into the bath. Notice that this result holds only when there are more than one channels for SB coupling ($\alpha\geq 2$), where the bath variables do not commute with each other ($[B^\alpha_{L_s+1},B^\beta_{L_s+1}]\neq 0$).   Throughout the paper, we focus on the ground state properties, while a similar mechanism has been proposed to explain the decoherence of the SPT edge modes in  quench dynamics\cite{McGinley2020}.

As an example of a stable SPT edge mode, the case with SB coupling $H_{sb}^{(c)}$ is considered, which preserve both the partial TRS and $Z_2\times Z_2$ SRS.  Fig.\ref{fig:fig2} (c) shows such SB coupling has almost no effect on the SPT edge mode on the site $L_s$, indicating that the SPT edge mode is protected by the $Z_2\times Z_2$ SRS. In principle, the second or higher order perturbations can be performed based on  $H_{sb}^{(c)}$. However, different from the TRS,  the $Z_2\times Z_2$ SRS is a unitary symmetry that does not change $i$ into $-i$, thus both the real and imaginary parts of $C_{L_s}^{\alpha\beta}$ in Eq.(\ref{eq:Hameff2}) preserve it and the corresponding effective Hamiltonian acts trivially on the SPT degenerate subspace.

\begin{figure}
\includegraphics[width=0.49\linewidth,bb=51 60 549 540]{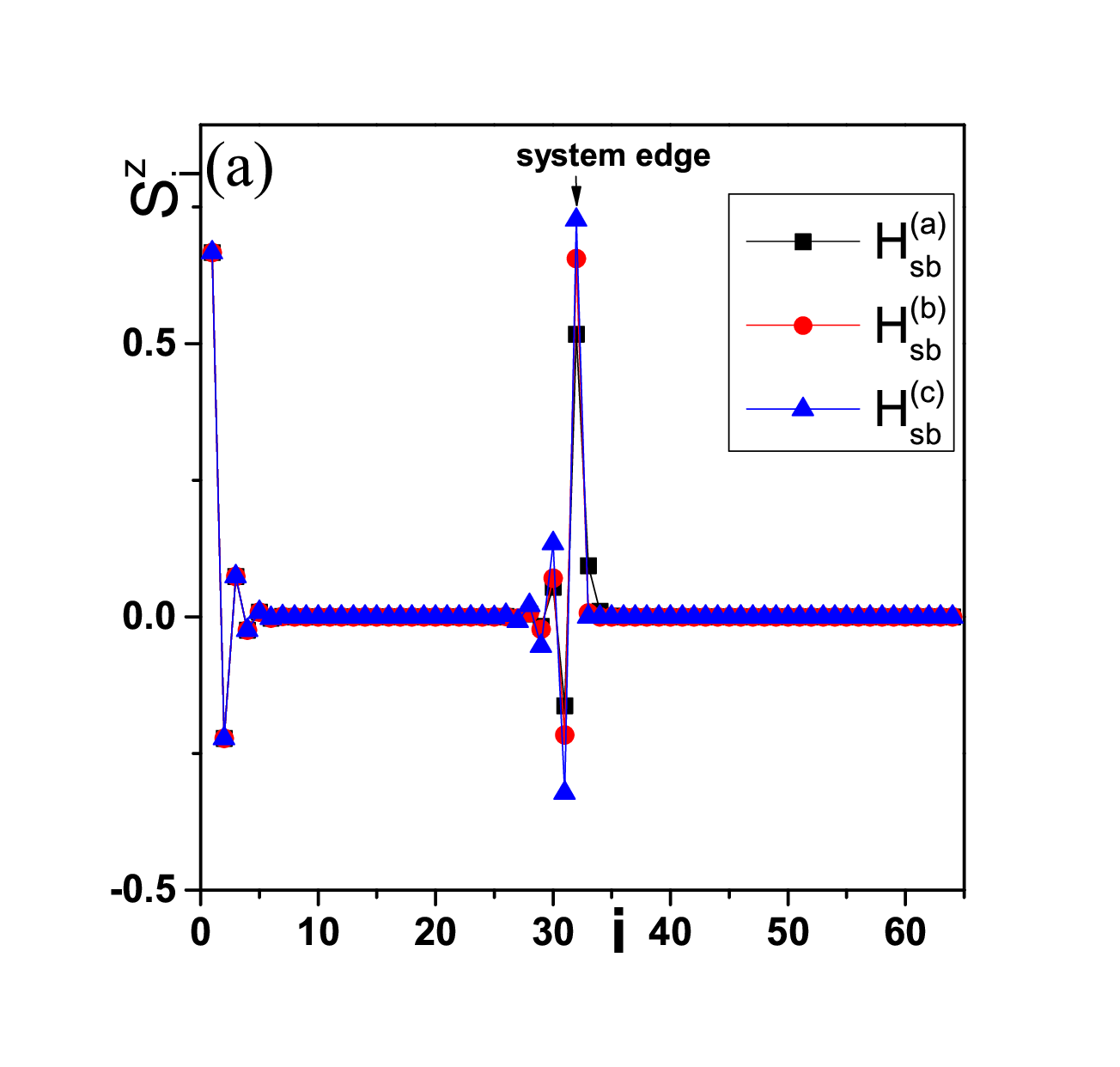}
\includegraphics[width=0.49\linewidth,bb=51 60 549 540]{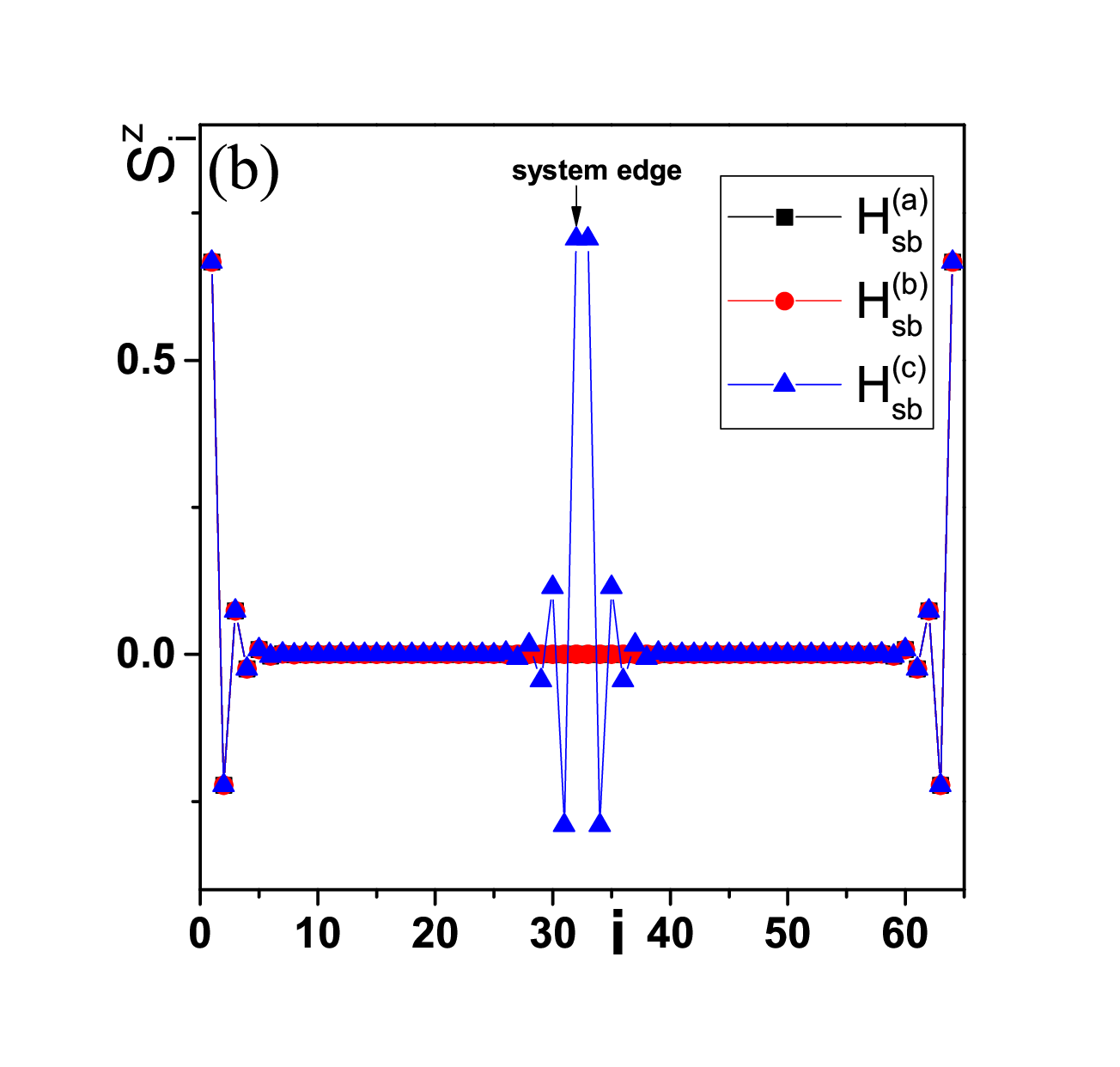}
\includegraphics[width=0.495\linewidth,bb=3 7 615 567]{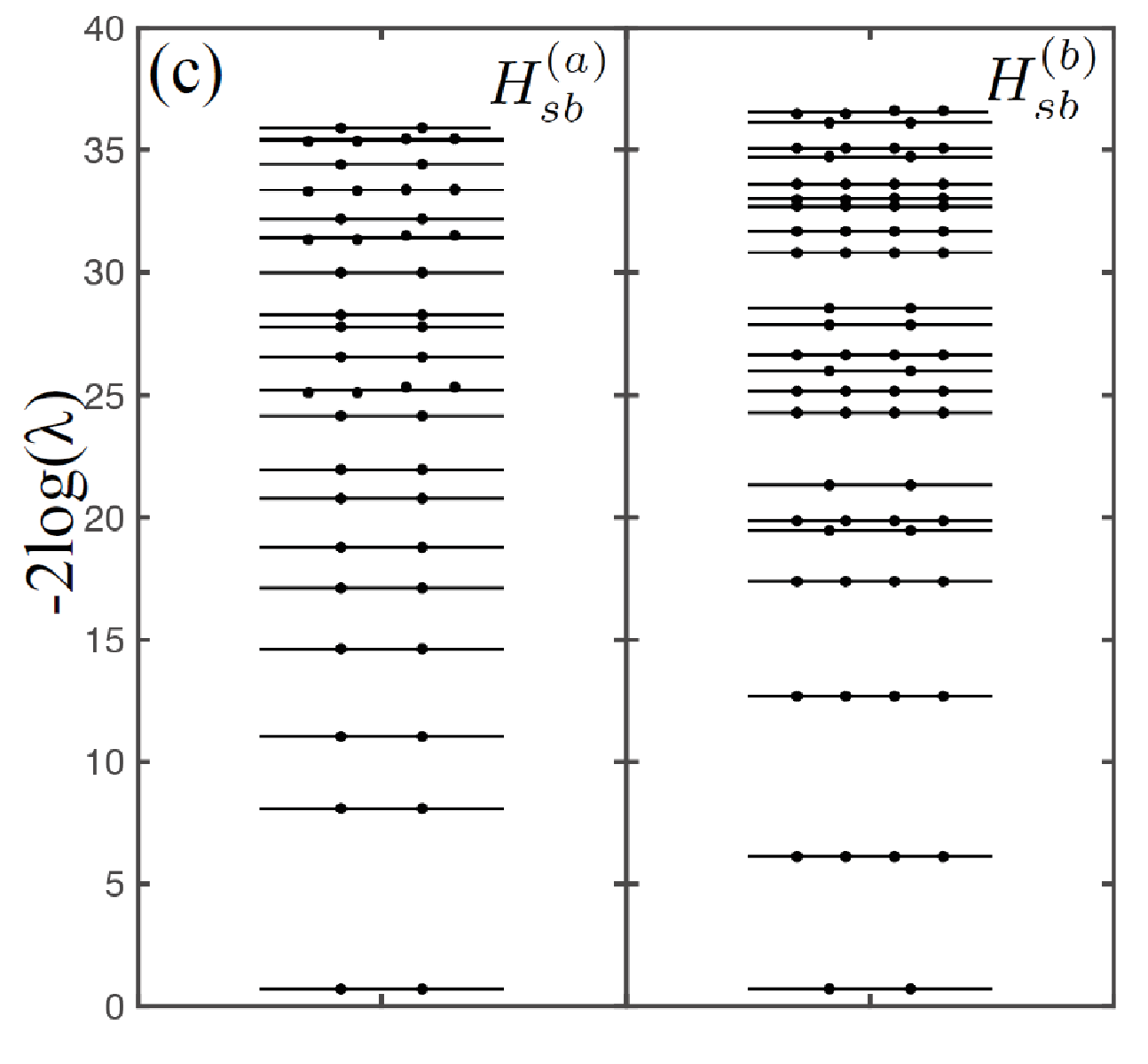}
\includegraphics[width=0.49\linewidth,bb=51 60 549 540]{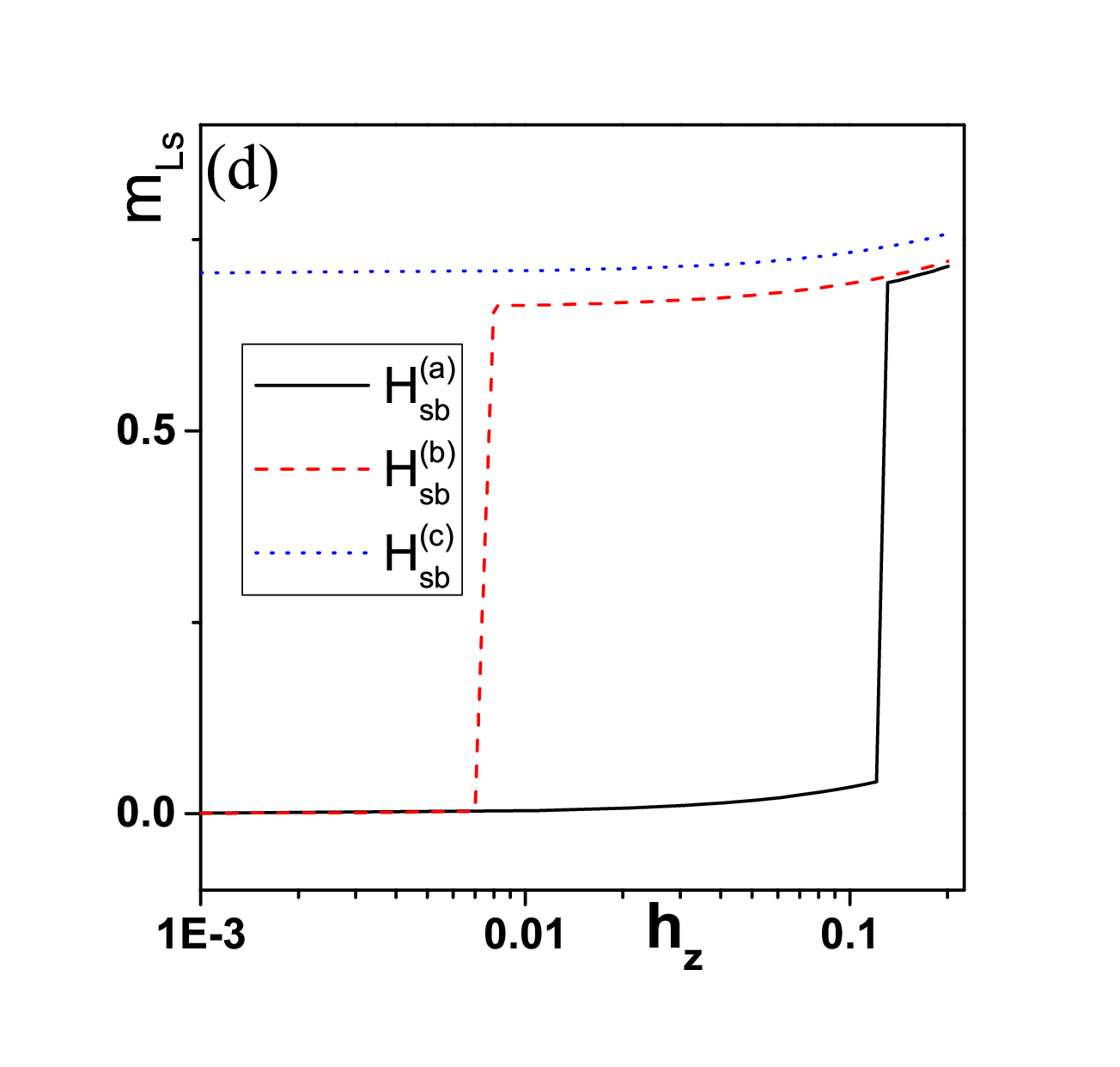}
\caption{(Color online).  The spatial distribution of the magnetization in a SPT system coupled to a (a) trivial gapped bath  and (b) topologically gapped bath via different SB couplings (square for $H_{sb}^{(a)}$, circle for $H_{sb}^{(b)}$, and triangle for $H_{sb}^{(c)}$). (c)Entanglement spectrum in the SB coupling bond  for the SB coupling $H_{sb}^{(a)}$ (left panel) and $H_{sb}^{(b)}$ (right panel). (d)Magnetization on the edge of the system $m_{L_s}$ as a function of the small local magnetic field $h_z$.  We choose  $D=0.5J$ for (a) and $L_s=L_b=32$, $J_b=0.1J$, and $J'=0.2J$ for (a)-(d). }\label{fig:fig3}
\end{figure}
{\it Quantum bath with a gap: topological versus non-topological --} In the gapless bath cases,  the partial symmetries of the SB coupling plays an important role in determining the SPT edge modes. Now we examine the role of  the bath itself. First, we consider a bath Ham:
\begin{equation}
\begin{small}
H_b=\sum_{i=L_s+1}^{L_b+L_s-1} [-J_b(S_i^xS_{i+1}^x+S_i^y S_{i+1}^y )+D (S_i^z)^2]\label{eq:Hame2}
\end{small}
\end{equation}
We focus on the case with $D\gg J_b$, where the bath is a non-topological gapped ground state (similar with the Mott insulator(MI) in the Bose-Hubbard model). Fig. \ref{fig:fig3} (a) shows that the SPT edge state is protected for all the SB couplings in Eq.(\ref{eq:Hama})-Eq.(\ref{eq:Hamc}). The non-topologically gapped bath plays a similar role as an open boundary, thus will not affect the edge state as long as the SB coupling preserves the global (system+bath) TRS symmetry. However, the edge mode vanishes  if the global TRS is broken  in the SB coupling ({\it e.g.} $S_{L_s}^x\otimes S_{L_s+1}^{xx}$\cite{Supplementary}).

 A different case is now considered, wherein the bath itself is also a gapped topological phase with SPT edge mode on its ends. The question to be answered is what type of SB coupling can couple the two edge modes on the ends of the system and bath chain.  To address this issue, we also choose the bath as a spin-1 AKLT chain:
\begin{equation}
\begin{small}
H_b=J_b\sum_{i=L_s+1}^{L_b+L_s-1} [\mathbf{S}_i\cdot \mathbf{S}_{i+1}+\frac 13 (\mathbf{S}_i\cdot \mathbf{S}_{i+1})^2]\label{eq:Hame3}
\end{small}
\end{equation}
the three SB couplings in Eq.(\ref{eq:Hama})-Eq.(\ref{eq:Hamc}) are still considered.  Fig.\ref{fig:fig3} (b) shows that for the SB couplings  $H_{sb}^{(a)}$ and $H_{sb}^{(b)}$, the edge mode at the end of the system chain is not protected by the TRS, while for $H_{sb}^{(c)}$, it is protected by the $Z_2\times Z_2$ SRS. These results are similar to the case with gapless bath, the mechanism is also similar. Take the $H_{sb}^{(b)}$ for an example, according to Eq.(\ref{eq:Hameff2}),  the effective coupling between the two edge spin-$\frac 12$s reads
\begin{small}
\begin{equation}
H_{sb}^{(2)}\sim\frac{J'^2}{\Delta} (\sigma_{L_s}^x\otimes \sigma_{L_s+1}^x+\sigma_{L_s}^y\otimes \sigma_{L_s+1}^y+\sigma_{L_s}^z\otimes\sigma_{L_s+1}^z)\label{eq:SBaklt}
\end{equation}
\end{small}
where $\sigma_{L_s}^{x,y,z}$ ( $\sigma_{L_s+1}^{x,y,z}$) are the $2\times2$ Pauli matrices  operating on the degenerate subspace spanned by the spin-$\frac 12$ edge mode on the ends of the system (bath) chain. Once these two spin-$\frac 12$ edge modes  are coupled to each other via the effective SB coupling Ham.(\ref{eq:SBaklt}) to form a spin singlet,  the edge spin-$\frac 12$ is not free, therefore is no longer protected. As a consequence, the composite system behaves similar to a whole AKLT chain where the bond connecting the system and bath is not qualitatively different from other bonds even though its original Hamiltonian ($H_{sb}^{(a)}$ and $H_{sb}^{(b)}$) significantly differs from the AKLT Hamiltonian in other bonds.   To verify this point numerically, the entanglement spectrum in the  bond $[L_s L_s+1]$ is plotted in Fig.\ref{fig:fig3} (c), which exhibits a even multiplicity of Schmidt values,  a signature of the Haldane phase\cite{Pollmann2010}. The singlet in the bond $[L_s L_s+1]$ can be verified by imposing a small magnetic field $H_h=-h_z(S_{L_s}^z+S_{L_s+1}^z)$.  Fig.\ref{fig:fig3} (d) shows that in the absence of singlet ($H_{sb}^{(c)}$), the edge spins are polarized by  infinitesimal magnetic field, in contrast, in the other two cases ($H_{sb}^{(a)}$ and $H_{sb}^{(b)}$), spin polarization can only occur when the external magnetic field is sufficiently large to overcome the energy gap between the spin singlet and triplet in this bond. In the case of $H_{sb}^{(a)}$, the threshold for such a transition is much larger than that in $H_{sb}^{(b)}$, which agrees with our perturbation analysis. The effective SB coupling in the case of $H_{sb}^{(b)}$ originates from the 2nd order perturbation, thus is considerably weaker than the direct coupling in $H_{sb}^{(a)}$.

{\it Discussion --} All the above-mentioned SB couplings preserve the global ($Z_2\times Z_2$ and TRS) symmetry.  It is shown that the SPT edge mode disappears in the case with a SB coupling breaking the global TRS\cite{Supplementary}, and it is interesting to explore the cases with other global symmetries breaking. Another important quantity that has not been explored so far is the string order, which is used to characterized the topological phases in the closed AKLT model. Our DMRG calculations show the string order correlations in the composite system (AKLT+XX) exhibit qualitatively different behaviors between the cases $H_{sb}^{(a)/(b)}$, and $H_{sb}^{(c)}$\cite{Supplementary}. This agrees with our conclusion based on the SPT edge modes that the $Z_2\times Z_2$ symmetry rather than TRS plays an important role in protecting the topological phase in the open quantum systems.

\begin{table}[htb]
\begin{tabular}{|p{2.5cm}<{\centering}|p{1.9cm}<{\centering}|p{1.9cm}<{\centering}|p{1.9cm}<{\centering}|}
\hline
 & $H_{sb}^{(a)}$/TRS:$\times$ $Z_2\times Z_2$:$\times$ & $H_{sb}^{(b)}$/TRS$:\surd$ $Z_2\times Z_2$:$\times$ & $H_{sb}^{(c)}$/TRS:$\surd$ $Z_2\times Z_2$:$\surd$  \\ \hline
gapless $H_b$\quad\quad(XX model) & $\times$ &  $\times$  &  $\surd$\\ \hline
 gapped  $H_b$ (non-topological)& $\surd$ & $\surd$ & $\surd$   \\ \hline
 gapped  $H_b$ (topological) & $\times$ & $\times$  &  $\surd$   \\ \hline
\end{tabular}
\caption{Dependence of the SPT edge mode on the partial symmetries of SB couplings and  features of the quantum bath.} \label{Table:1}
\end{table}

{\it Conclusion and outlook --} In conclusion, the edge mode of an SPT system upon coupling with a quantum bath is systemically studied. TRS is observed to plays a special role comparing to other unitary symmetries.  The major conclusion is summarized in Table.I. The main finding is that even though the ground state degeneracy for a composite quantum system can be protected by global TRS, the partial TRS only with respect to the system variables is not sufficient to prevent the original SPT edge mode from diffusing into the quantum bath, which makes it useless for quantum computation.

Even though these results are obtained based on a specific model of a $S=1$ quantum spin chain,  these conclusions are believed to be general in the sense that they can be applied to other open quantum systems({\it e.g.} the spin-$\frac 12$ or fermionic SPT models) with various quantum bath ({\it e.g.} a bosonic bath the TRS of which is defined differently from that of spin systems). Future developments will include higher dimensional generalization of these results. For instance, for a TSR-protected $Z_2$ topological insulator, it would be interesting to analyze  how the gapless edge modes and the corresponding quantized Hall conductance are affected by the quantum bath\cite{Dyke2017,Deng2021}.

\begin{acknowledgements}
{\it Acknowledgments}.---ZC acknowledge helpful discussion with Zheng-Xin Liu. This work is supported by the National Key Research and Development Program of China (Grant No. 2020YFA0309000 and No. 2016YFA0302001), NSFC of  China (Grant No.11574200),  Shanghai Municipal Science and Technology Major Project (Grant No.2019SHZDZX01).
\end{acknowledgements}


\end{document}